\documentclass{eptcs}
\providecommand{\event}{DCM 2010} 
\usepackage{breakurl}             
\usepackage{latexsym} 
\usepackage{graphicx} 
\newcommand{\uj}{\newcommand} 
\uj{\vsp}{\vspace{1ex}} 
\uj{\vspp}{\vspace{.5ex}}
\uj{\vspe}{\vsp \newline} 
\uj{\vspm}{\vspace*{-.7cm}} 
\uj{\vspmm}{\vspace*{-.6cm}} 
\uj{\vspmmm}{\vspace*{-2mm}} 
\uj{\vspmini}{\vspace*{-1mm}}
\uj{\vspminni}{\vspace*{-.6mm}}
\uj{\nyil}{\rightarrow} 
\uj{\nnyil}{\Rightarrow}
\uj{\paral}{\parallel} 
\uj{\set}{\mbox{\bf Set}} 
\uj{\rel}{\mbox{\bf Rel}}
\uj{\egy}{\mbox {{\bf 1}}}
\uj{\eps}{\epsilon }
\uj{\calm}{\mathcal{M}}
\uj{\kapo}{\updownarrow}
\uj{\calp}{\mathcal{P}}
\uj{\calt}{\mathcal{T}}
\uj{\calc}{\mathcal{C}}
\uj{\calg}{\mathcal{G}}
\uj{\calf}{\mathcal{F}}
\uj{\cala}{\mathcal{A}}
\uj{\calgsig}{\calg (\Sigma )}
\uj{\tot}{{\bf T}}
\uj{\ibf}{{\bf i}}
\uj{\jbf}{{\bf j}}
\uj{\wbf}{{\bf w}}
\uj{\robf}{\rho }
\uj{\biset}{\mbox {\bf Bset}}
\uj{\ima}{\mbox {\bf IMA}}
\uj{\sdcc}{\mbox {\bf SDCC}}
\uj{\bmc}{{\bf c}}
\uj{\imm}{\mathcal {I}}
\uj{\alp}{\mathcal {A}}
\uj{\sdc}{\mathcal {S}}
\uj{\dil}{{\em -dil}}
\uj{\bind}{\mbox{\bf Ind}}
\uj{\cali}{\mathcal{I}}
\uj{\vege}{\hspace{1em} \raisebox{-.5ex}{{\large $\Box $}}} 
\newtheorem{theorem}{Theorem}
\newtheorem{definition}{Definition}
\title{Turing Automata and Graph Machines\thanks{Work partially supported by Natural
Science and Engineering Research Council of Canada, Discovery Grant \#170493-03.}}
\author{Mikl\'os Bartha
\institute{Deprtment of Computer Science\\
Memorial University of Newfoundland\\
St.\ John's, NL, Canada}
\email{bartha@mun.ca}
}
\def\titlerunning{Turing Automata}
\def\authorrunning{M. Bartha}
\begin{document}
\maketitle

\begin{abstract}
Indexed monoidal algebras are introduced as an equivalent structure for
self-dual compact closed categories, and a coherence theorem is proved
for the category of such algebras. Turing automata and Turing graph machines 
are defined by generalizing the classical Turing machine concept, so that
the collection of such machines becomes an indexed monoidal algebra. 
On the analogy of the von Neumann data-flow computer architecture,
Turing graph machines are proposed as potentially reversible
low-level universal computational devices, and a truly reversible molecular 
size hardware model is presented as an example.
\end{abstract}
\section{Introduction}
The importance of reversibility in computation has been argued at several
platforms in connection with the speed and efficiency of modern-day computers.
As stated originally by Landauer \cite{landa} and re-emphasized by Abramsky \cite{abr}:
``it is only the logically irreversible operations in a physical computer that
necessarily dissipate energy by generating a corresponding amount of entropy
for every bit of information that gets irreversibly erased''. Abramsky's
remedy for this situation in \cite{abr} is to translate high level functional
programs in a syntax directed way into a simple kind of automata which are
immediately seen to be reversible. The concept strong compact closed 
category \cite{abst} has been introduced and advocated as a theoretical 
foundation for this type of reversibility.

The problem of reversibility, however, does not manifest itself at the software
level. Even if we manage to perform our programs in reverse, it is not
guaranteed that information will not be lost during the concrete physical
computation process. To the contrary, it may get lost twice, once in each
direction. The solution must therefore be found at the lowest hardware level.
Our model of Turing graph machines is being presented as a possible hardware
solution for the problem of reversibility, but follows Abramsky's structural
approach. We even go one step further by showing how computations can be done
in a virtually undirected fashion under the theoretical umbrella of self-dual
compact closed categories. In practical terms we mean that, unlike in 
synchronous systems (e.g.\ sequential circuits), where the information is
propagated through the interconnections (wires) between the functional
elements (logical gates) always in the same direction, in a Turing graph
machine the flow of information along these interconnections takes a 
direction that is determined dynamically by the current input and state of the machine.  
We are going to reconsider self-dual compact closed categories as
indexed monoidal algebras and prove a coherence theorem to establish undirected
graphs -- constituting the basic underlying structure for Turing graph machines -- as
free indexed monoidal algebras generated by the ranked alphabet consisting of
the star graphs.

Different parts of this paper need not be read in a strict sequential order. 
In-depth knowledge of algebra and category theory is only required in Section~2
and Section~3. The reader less familiar with categories could still understand
the concept of Turing automata and Turing graph machines in Section~5, and appreciate the 
main contribution of this work. The paper, being a short summary of rather
complex theoretical results, admittedly elaborates only on those
connections to these results that are directly related to their presentation.
One may, however, recognize structures familiar from linear logic,
game semantics, communicating concurrent processes, iteration theories, interaction nets, 
and the Geometry of Interaction program in general. These connections will be spelled
out in a future extended version of the present summary.
\section{Self-dual compact closed categories}
In this section we shall assume familiarity with the concept of {\em
symmetric monoidal categories\/} \cite{mcl}. Even though our main concern is with
strict monoidal categories, the algebraic constructions presented in Section~3
can easily be adjusted to cover the general case. From this point on,
unless otherwise stated, by a monoidal category we shall always mean a strict 
symmetric one. 

Let $\calc $ be a monoidal category with tensor $\otimes $ and unit object $I$.
Recall from \cite{cc,tra} that $\calc $ is {\em compact closed\/} if every
object $A$ has a left adjoint $A^*$ in the sense that there exist morphisms
$d_A:I\nyil A\otimes A^*$ (the unit map) and $e_A:A^*\otimes A\nyil I$ (the
counit map) for which the two composites below result in the identity
morphisms $1_A$ and $1_{A^*}$, respectively.
 \[ A=I\otimes A\nyil _{d _A\otimes  1_A}(A\otimes A^*)
\otimes A=A\otimes (A^*\otimes A)\nyil _{1_A\otimes e_A}
A\otimes I=A, \vspmmm\] 
\[ A^*=A^*\otimes I\nyil _{1_{A^*} \otimes d_A}A^*\otimes 
(A\otimes A^*)=(A^*\otimes A)\otimes A^*\nyil _{e_A\otimes 1_{A^*}}
I\otimes A^*=A^*. \vspminni \]
By virtue of the adjunctions $A\dashv A^*$ there is a natural isomorphism between the 
hom-sets $\calc (B\otimes A,C)$ and $\calc (B,C\otimes A^*)$ for every objects $B,C$,
hence the name ``compact closed'' category. Category $\calc $ is {\em
self-dual\/} compact closed (SDCC, for short) if $A=A^*$ for each object $A$.
The category $\sdcc $ has as objects all locally small \cite{mcl}
SDCC categories, and as
morphisms monoidal functors preserving the given self-adjunctions.

A well-known SDCC category (not strict, though) is the category $(\rel ,\times)$ of 
small sets and
relations with tensor being the cartesian product $\times $.
We shall only use this category as an example to explain the idea of indexing on it.
Recall from \cite {bur,hel} that an {\em indexed family of sets\/} is simply a functor
$\cali : \bind\nyil \set $, where $\bind $ is the index category.  In our example,
$\bind $ is the monoidal category $(\set ,\times)$ as a subcategory of $(\rel ,\times )$
and $\cali $ is the covariant powerset functor $\calp $, which is of course not
monoidal. Relations $A\nyil B$ are, however, still subsets of $A\times B$, and as 
such they can be indexed by morphisms (functions) $A\times B\nyil C$ in $\set $.
For any two objects (sets) one can then consider the binary operation $\oplus _{A,B}
(=\times ): \calp (A)\times \calp (B)\nyil \calp (A\times B)$, and the unary operation
trace, $\kapo _{A,B}:\calp (A\times A\times B)\nyil \calp (B)$ for which $b\in \kapo _{A,B}R$
iff $\exists a\in A\; (a,a,b)\in R$. The concept indexed monoidal algebra arises from
observing the equational algebraic laws satisfied by these operations and their 
relationship to indexing. 

  Regarding the index monoidal category $\bind $, one would like to have it as narrow as possible.
The best choice would be the collection of permutations in $\bind $, which, unfortunately,
fails to be a subcategory in general. To get around this 
problem we shall introduce so called permutation symbols as unary operations,
which will be responsible for the task of indexing in a coherent way.
\section{Indexed monoidal algebras}
In this section we introduce the category $\ima $ of indexed monoidal algebras
along the lines of the pioneer work \cite{hel}, and establish 
an equivalence between the categories $\ima $ and $\sdcc $.

Let $S$ be a set of abstract sorts, and consider the free monoid $S^*$
generated by $S$.  For a word (string) $w\in S^*$, $|w|$ will denote the
length of $w$ and () will stand for the empty string. By an 
$S$-{\em permutation\/} we mean a pair $(w,\pi )$, where $w=s_1\ldots s_n$ is
a string with $|w|=n\geq 0$, and $\pi $ is a permutation $n\nyil n$. We shall
use the notation $w\sharp \pi $ for $(w,\pi)$, and say that $w\sharp \pi $ is an
$S$-permutation $w\nyil \pi (w)$, where $\pi (w)$ is the string $s_{\pi
(1)}\ldots  s_{\pi (n)}$.  If $v$ and $w$ are strings of length $n$ and $m$,
respectively, then  $c_{v,w}$ will denote the $S$-permutation $vw\sharp
x_{n,m}$ in which $x_{n,m}$ is the block transposition $n+m\nyil m+n$.

The collection of $S$-permutations can naturally be equipped with the operations
composition ($\bullet $) and tensor ($\oslash $), which structure, together with the 
identities $1_w=w\sharp id _n$ and symmetries $c_{v,w}$, defines a monoidal 
category $\Pi _S$ over the set of
objects $S^*$. See e.g.\ \cite [Definition~1]{acta} covering the single-sorted
case. The category $\Pi_S$ is $S$-initial in the sense that, for every monoidal category
$\calc $ and mapping $\chi $ from $S$ to the objects of $\calc $, there exists a unique monoidal
functor $\chi :\Pi_S\nyil \calc $ extending $\chi $ on objects. See again \cite
[Corollary 1]{acta}  for a proof in the single-sorted case. 

Now let $(M,I,\otimes)$ be a cancellative monoid, fixed for the rest of the paper.
Since the elements of $M$ are meant to be objects in an appropriate monoidal
category, they will be denoted by capital letters. With a slight abuse of the
notation, $M^*$ will no longer mean the free monoid generated by $M$, rather, its
quotient by the equation $I=()$. Accordingly, by $\Pi _M$ we mean the monoidal
category of $M$-permutation symbols, rather than that of ordinary $M$-permutations.
We do so in order to accommodate the assumption that our monoidal categories
are strict. Permutations over $M$ in this new sense will then be called 
$M$-permutation {\em symbols\/} to restore unambiguity. Let $\eps _M:M^*\nyil M$
be the unique homomorphism (counit map) determined by the identity function on $M$.
Again, this time with a heavier abuse of the terminology and coherence, the $M$-permutation
symbol $w\sharp \pi :w\nyil \pi (w)$ will also be called one with ``domain'' $A=\eps _M(w)$
and ``codomain'' $B=\eps _M(\pi (w))$. As an escape, however, we shall use the distinctive
notation $w\sharp \pi :A\nnyil B$ and say that permutation symbols $\rho _1:A\nnyil B$
and $\rho _2:B\nnyil C$ are {\em composable\/} if they are such as proper morphisms
in the category $\Pi_M$. 

Let $\robf =\wbf \sharp \alpha $ be an $M^*$-permutation symbol with $\wbf =
u_1\ldots u_n$, where $u_i=A_{i,1}\ldots A_{i,m_i}$. In the monoidal category
$\Pi _{M}$, $\robf $ defines an $M$-permutation symbol
$\epsilon _{M}(\robf ): u_1\ldots u_n\nyil u_{\alpha (1)}\ldots u_{\alpha (n)}$.
On the other hand, $\robf $ also gives rise naturally to the $M$-permutation
symbol $\robf /\epsilon _{M}: B_1\ldots B_n\nyil B_{\alpha (1)}\ldots 
B_{\alpha (n)}$, where $B_i=\otimes _j A_{i,j}$. Clearly, 
$\epsilon _{M}(\robf)$ and $\robf /\epsilon _M$ define the same permutation
$\otimes _i B_i\nyil \otimes _i B_{\alpha (i)}$ in every monoidal category
having $(M,I,\otimes )$ as its object structure. Therefore we say that these
two $M$-permutation symbols $\otimes _i B_i\nnyil \otimes _i B_{\alpha (i)}$
are {\em equivalent\/} and write  $\epsilon _{M}(\robf)\equiv 
\robf /\epsilon _M$. As a trivial, but representative example:
$1_A\oslash 1_B = 1_{AB}\equiv 1_{A\otimes B}$.

 We shall be dealing with $M$-sorted algebras $\calm =\{\calm _A\,|\,A\in M\}$ having
the following operations and constants.
\vsp \newline 
  -- For each $M$-permutation symbol $\rho:A\nnyil B$, a unary operation 
    $\rho :\calm _A\nyil \calm _B$.
\newline
  -- For each $A,B\in M$, a binary operation sum, $\oplus :\calm _A\times\calm _B\nyil
     \calm _{A\otimes B}$.
\newline
  -- For each $A\in M$, a constant $\egy _A\in \calm _{A\otimes A}$.
\newline
  -- For each $A,B\in M$, a unary operation trace, $\kapo _A:\calm _{A\otimes A\otimes B}
     \nyil \calm _B$.
\vsp

To emphasize the categorical nature of such algebras we call the elements $f\in \calm _A$
morphisms and write $f:A$. We also write $f:A\nyil B$ as an alternative
for $f:A\otimes B$.   
Note that cancellativity of $M$ is required in order to make the trace operation sound.
Moreover, the accurate notation for trace would be $\kapo _{A,B}$, but the intended object 
$B$ will always be clear from the context.
Also notice the boldface notation $\egy _A:A\nyil A$ as opposed to $1_w:w\nyil w$.
For better readability we shall write $f\cdot \rho $ for $f\rho $, that is, for indexing
$f$ by permutation symbol $\rho $.

Composition ($\circ $) and tensor ($\otimes $) are introduced in $\calm $
as derived operations in the following way.      
\vsp\newline
  -- For $f:A\nyil B$ and $g:B\nyil C$, $f\circ g=\kapo _B((f\oplus g)\cdot 
    (c_{A,BB}\oslash 1_C))$. 
\newline
  -- For $f:A\nyil B$ and $g:C\nyil D$, $f\otimes g=(f\oplus g)\cdot (1_A\oslash 
     c_{B,C}\oslash 1_D)$.
\vsp\newline
See Fig.\ 1. Again, the accurate notation for $\circ $ and $\otimes $ would use the 
objects $A,B,C,D$ as subscripts, but these objects will always be clear 
from the context. Observe that the above definition of composition and tensor
is in line with the traced monoidal category axioms in \cite{tra,concur}. 
Regarding composition, see also \cite[Identity $X_3$]{acta}. As we shall point out
in Theorem~1 below, our trace operation models the so called 
``canonical'' trace concept (cf.\ \cite{tra}) in SDCC categories. 
\begin{figure}[h]
\begin{center}
\includegraphics[scale=0.8]{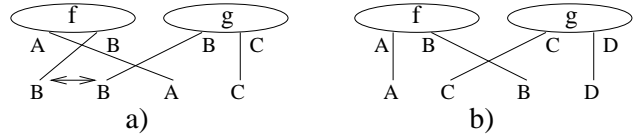}
\end{center}
\vspmm
\caption{Composition (a) and tensor (b) in $\calm $}
\vspmmm \vspmini
\end{figure}
\begin{definition}
{\em An {\em indexed monoidal algebra\/} over $M$ is an $M$-sorted algebra $\calm =
\{\calm _A\,|\,A\in M\}$ equipped with the operations and constants listed above,
which satisfies the following equational axioms.
\vsp\newline
  I1.\ {\em Functoriality of indexing}

   $f\cdot(\rho_1\bullet \rho _2)=(f\cdot \rho _1)\cdot \rho_2$ for $f:A$ and
   composable $\rho_1:A\nnyil B$, $\rho _2:B\nnyil C$;

   $f\cdot 1_{A}=f$ for $f:A$.
\newline
 I2.\ {\em Naturality of indexing}

   $(f\oplus g)\cdot (\rho _1\oslash \rho _2)=f\cdot \rho _1\oplus g\cdot \rho_2$
    for $f:A$, $g:B$, $\rho _1:A\nnyil C$, $\rho _2:B\nnyil D$;

   $(\kapo _A f)\cdot \rho =\kapo _A(f\cdot (1_{AA}\oslash \rho ))$
    for $f:A\otimes A\otimes B$, $\rho :B\nnyil C$.
\newline
I3.\ {\em Coherence}

    $f\cdot \rho _1 =f\cdot \rho _2 $ for $f:A\nyil B$, whenever $\rho_1\equiv \rho _2$.
\newline
 I4.\ {\em Associativity and commutativity of sum}

    $(f\oplus g)\oplus h=f\oplus (g\oplus h)$ for $f:A$, $g:B$, $h:C$;

    $f\oplus g=(g\oplus f)\cdot c_{A,B}$ for $f:A$, $g:B$.
\newline
 I5.\ {\em Right identity}

    $f\circ \egy _B=f$ and $f\oplus \egy _I =f$ for $f:A\nyil B$.
\newline
 I6.\ {\em Symmetry of identity}

    $\egy _A\cdot c_{A,A}=\egy _A$.
\newline
 I7.\ {\em Vanishing} 

    $\kapo _I f=f$ for $f:A$;

    $\kapo _{A\otimes B}f=\kapo _B(\kapo _A f\cdot (1_A\oslash c_{B,A}\oslash 1_{BC}))$
     for $f:A\otimes B\otimes A\otimes B\otimes C$.
\newline
 I8.\ {\em Superposing}

    $\kapo _A(f\oplus g)=\kapo _A f\oplus g$ for $f:A\otimes A\otimes B$, $g:C$.
\newline
 I9.\ {\em Trace swapping}

    $\kapo _B(\kapo _Af)=\kapo _A(\kapo _B(f\cdot (c_{AA,BB}\oslash 1_C)))$
    for $f:A\otimes A\otimes B\otimes B\otimes C$.
}
\end{definition}
The algebra $\calm $ is called {\em small\/} if $M$ is a small monoid 
in the sense of \cite{mcl} and the
sets $\calm _A$ are also small for every $A\in M$.

  Let $\calm $ and $\calm '$ be indexed monoidal algebras. An {\em indexed monoidal 
homomorphism\/} $F:\calm \nyil \calm '$ is a pair $(h, \{ F_A\,|\,A\in M\})$, where
$h$ is a monoid homomorphism $M\nyil M'$
and $F_A: \calm _A\nyil \calm '_{h(A)}$ are mappings
that determine a homomorphism in the usual algebraic sense. With respect to indexing
we mean that for every $f:A$ and $\rho :A\nnyil B$, $F_B(f\cdot \rho)=(F_A f)\cdot 
h^*\rho $, where $h^*$ is the unique monoidal functor $\Pi _M\nyil \Pi _{M'}$ determined
by $h$. The category $\ima $ then consists of all small indexed monoidal
algebras as objects and indexed monoidal homomorphisms as morphisms.
\begin{theorem}
The categories $\ima $ and $\sdcc $ are equivalent.
\end{theorem}
{\em Proof.\ \ }
Let $\calm $ be a small indexed monoidal algebra over $M$, and define the monoidal category
$\calc =\sdc \calm $ over the objects $M$ as follows.
Morphisms $A\nyil B$ and identities in $\calc $ are exactly those in $\calm $, while
composition and tensor are adopted from $\calm $ as derived operations.
Symmetries $\bmc _{A,B}: A\otimes B\nyil B\otimes A$ in $\calc $ are the morphisms 
$\egy _{A\otimes B}\cdot (1_{AB}\oslash c_{A,B})$.
In general, every permutation symbol $\rho :w\nyil w'$ is represented in $\sdc \calm $
as $\egy _{\eps _M(w)}\cdot (1_w\oslash \rho ):\eps _M(w)\nyil \eps _M(w')$. 
For each self-adjunction $A \dashv A$, the unit
map $d_A:I\nyil A\otimes A$ and the counit map $e_A:A\otimes A\nyil I$ are both the 
identity $\egy _A:A\otimes A$. It is essentially routine to check that $\sdc \calm $
is a locally small SDCC category. Below we present the justification of some milestone
equations, which can easily be developed into a complete rigorous proof.
\vspe
1.\ \ {\em Symmetry of trace, and canonical trace}
\vsp

    $\kapo _A f=\kapo _A (f\cdot (c_{A,A}\oslash 1_B))=
     \egy _A\circ _{(A\otimes A)}f$ for $f:A\otimes A\otimes B$.
\vspp\newline
See Fig.\ 2.
\begin{figure}[h]
\begin{center}
\includegraphics[scale=0.55]{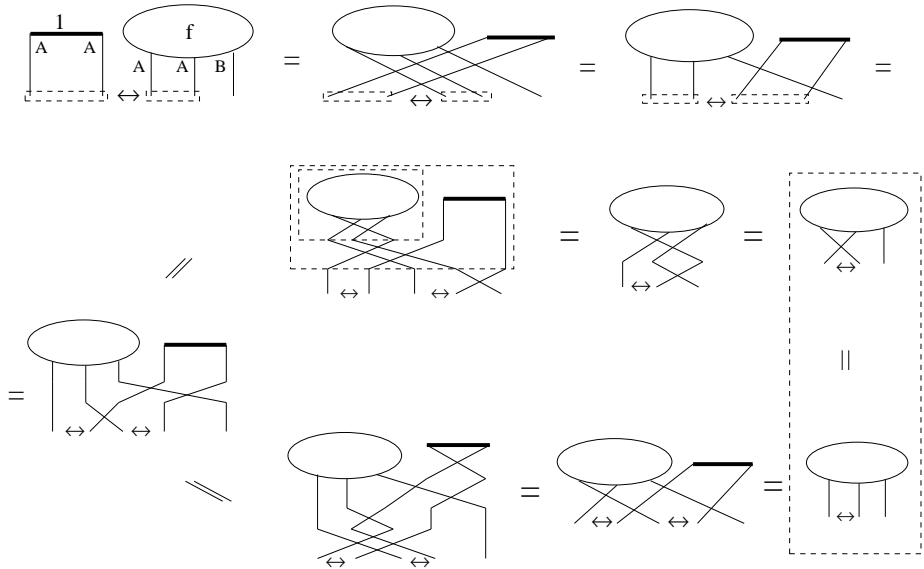}
\end{center}
\vspmini
\caption{Symmetry of trace, and canonical trace (read from right to left, bottom-up)}
\end{figure}
\newline
2.\ \ {\em Left identity}
\vsp
 
     $\egy _A\circ f=f$ for $f:A\nyil B$
\vspp\newline
See Fig.\ 3.
\begin{figure}
\begin{center}
\includegraphics[scale=0.6]{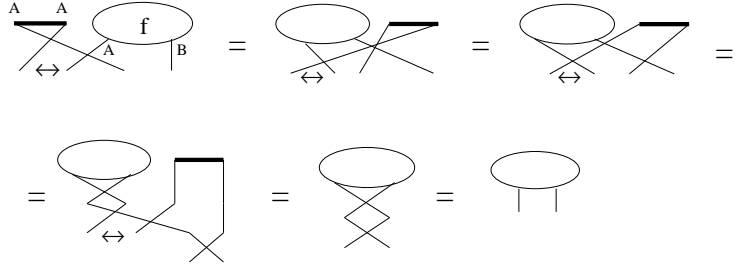}
\end{center}
\vspmini
\caption{Left identity}
\end{figure}
\newline
Notice that the symmetry of trace and that of $\egy _A$ have both been used in
the proof.
\vspe
3.\ \ {\em Tensor of identity}
\vsp

   $\egy _{A\otimes B}=\egy _A\otimes \egy _B$
\vspp\newline
See Fig.\ 4.
\begin{figure}
\begin{center}
\includegraphics[scale=0.53]{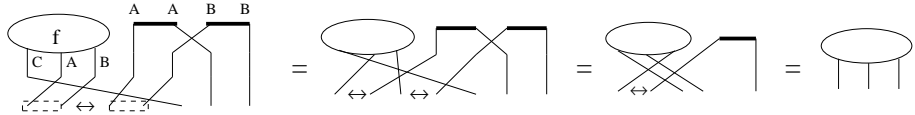}
\end{center}
\vspmini
\caption{Tensor of identity, take $f=\egy _{A\otimes B}$}
\vspmini 
\end{figure}
\vspe
The definition of functor $\sdc $ on (homo-)morphisms is evident, and left to the reader.

Conversely, let $\calc $ be a locally small SDCC category over $M$ as objects, and define the indexed
monoidal algebra $\calm =\imm \calc $ as follows. For each $A\in M$, $\calm _A=
\calc (I,A)$, the (small) set of morphisms $I\nyil A$ in $\calc $. Since $\calc $
is symmetric, every permutation symbol $\rho :A\nnyil B$ determines a permutation
$\rho _{\calc }:A\nyil B$ in $\calc $. Then, for $f:A$,
define $f\cdot \rho =f\circ \rho _{\calc }$. Notice that indexing
indeed becomes the restriction of the covariant $hom$ functor to permutations, as intended.
For $f:A$ and $g:B$, $f\oplus _{\calm} g=f\otimes _{\calc} g: I\nyil A\otimes B$ and
$\egy _A=d_A:I\nyil A\otimes A$. For $f:A\otimes A\otimes B$, $\kapo _A f$ is 
defined as the canonical trace of the morphism $f _A:A\nyil A\otimes B$ in
$\calc $ that corresponds to $f$ according to compact closure. That is,
$\kapo _A f$ is the morphism $f\circ (e _A\otimes 1_B):I\nyil B$ in
$\calc $.

In the light of this translation, each of
the equations I1-I9 is either a standard monoidal category axiom or has been observed
in \cite{tra,cc} for traced monoidal or compact closed categories.
Thus, $\calm $ is an indexed monoidal algebra. The specification of
functor $\imm $ on morphisms (monoidal functors) is again straightforward.

By definition, $\imm (\sdc \calm )=\calm$. On the other hand, the only difference
between the monoidal categories $\calc $ and $\sdc (\imm \calc )$ is that the hom-sets 
$A\nyil B$ in the latter are identified with the ones $I\nyil A\otimes B$ of the former,
using the natural isomorphisms given by the self-adjunctions $A\dashv A$. 
In other words, morphisms $A\nyil B$ in $\sdc (\imm \calc )$ -- as provided for by compact
closure -- are simply renamed as they appear in $\calc (I,A\otimes B)$. Thus,
there exists a natural isomorphism between the functors $1_{\sdcc }$ and $\sdc \imm $,
so that the categories $\ima $ and $\sdcc $ are equivalent as stated. 
\vege 
\section{Coherence in indexed monoidal algebras}
In general, a coherence result for some type $\tau $ of monoidal categories is about
establishing a left-adjoint for a forgetful functor from the category $\tot $ of
$\tau $-monoidal categories into an appropriate syntactical category, and providing a graphical
characterization of the free monoidal $\tau $-categories so obtained. For some typical
examples, see \cite{mcl,cc,acta,tcs}. In this section we present such a 
coherence result for SDCC categories, but phrase it in terms of indexed monoidal algebras.
The graphical language arising from this result will justify our efforts in the previous
section to reconsider SDCC categories in the given algebraic context.

For a set $S$ of sorts, an $S$-{\em ranked alphabet} (signature) is a set 
$\Sigma =\cup (\Sigma _w\,|\,w\in S^*)$, where $\Sigma _v\cap \Sigma _w=\emptyset $
if $v\neq w$. A morphism $\Omega :\Sigma \nyil \Delta $ between ranked alphabets of sort
$S$ and $T$, respectively, is an {\em alphabet mapping\/} consisting of a function
$\omega :S\nyil T$ and a family of mappings $\Omega _w:\Sigma _w\nyil \Delta _{\omega (w)}$.
(The unique extension of $\omega $ to strings is denoted by $\omega $ as well.)
Every indexed monoidal algebra $\calm =\{\calm _A\,|\,A\in M\}$ can be considered
as an $M$-ranked alphabet $\Sigma =\alp \calm $ in such a way that $\Sigma _w=
\{ f_w\,|\,f:\eps _M(w) \mbox{\ in\ } \calm \}$ for every $w\in M^*$. (The identification
$I=()$ is still in effect for $M^*$.) We use a subscript to distinguish
between instances of $f$ belonging to different ranks. If $F=(h, \{ F_A\,|\,A\in M\}):
\calm \nyil \calm '$ is a homomorphism, then $\alp F:\alp \calm \nyil \alp \calm '$ is 
the alphabet mapping $(h, \{ H_w\,|\,w\in M^*\})$ for which $H_w(f_w)=F_{\eps _M(w)}(f)$.
Our aim is to provide a left adjoint for the functor $\alp $. In algebraic terms this 
amounts to constructing the indexed monoidal algebra freely generated by a given 
$S$-ranked alphabet $\Sigma $.

Let $\Sigma $ be an $S$-ranked alphabet. By a $\Sigma $-{\em graph\/} we mean a
finite undirected and labeled multigraph $G=(V,E,l)$ with vertices (nodes) $V$, edges $E$, and
labeling $l:V\nyil \Sigma \cup \{ in _A, \angle _A\,|\, A\in S\}$, where $in _A$ and
$\angle _A$ are special symbols not in $\Sigma $ with rank $A$ and (), respectively.
Vertices labeled by $in _A$ ($\angle _A$) will be called {\em interface\/} (respectively,
{\em loop\/}) vertices. All other vertices, as well as the edges connecting them will be 
called {\em internal\/}. It is required that the label of each node $u$ be consistent 
with its degree $d(u)$, so that if $l(u)\in \Sigma _w$, then $|w|= d(u)$. Each point at which 
an edge impinges on $u$ is assigned a serial number $1\leq i\leq n=d(u)$ and a sort
$A_i$ in such a way that $w=A_1\ldots A_n$. Adopting a terminology from \cite
{tcs,mil}, such points will be referred to as {\em ports}. Edges, too,
must be consistent with the labeling in the sense that each edge connects two
ports of the same sort, and each port is  an endpoint of exactly one edge. The
interface nodes themselves are assigned a serial number, so that one can speak
of a $\Sigma $-graph $G:w$ with $w$ being the string of sorts assigned to (the
unique ports of) the interface vertices in the given order. See Fig.~5a for an
example  $\Sigma $ graph $G:AB$, where $\Sigma =\Sigma _{BA}\cup \Sigma
_{ABA}$ with $\Sigma _{BA}=\{ f\}$ and $\Sigma _{ABA}=\{ g\}$. Symbols in
$\Sigma $ are represented as {\em atomic\/} $\Sigma $-graphs in the way
depicted by Fig.~5b. 
\begin{figure}[h]
\begin{center} 
\includegraphics[scale=0.7]{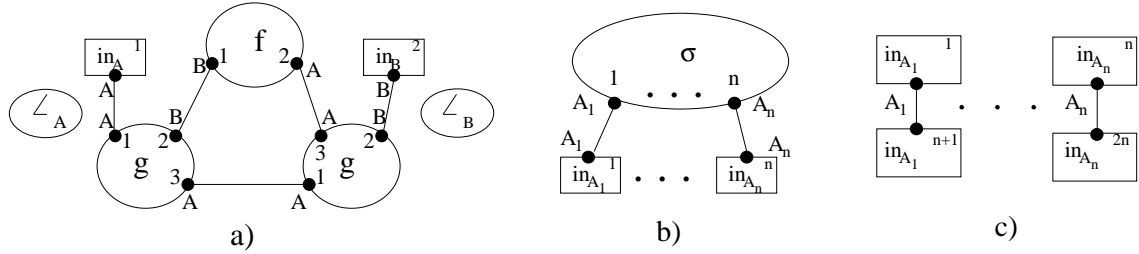}
\end{center}
\vspmini
\caption{$\Sigma $-graphs}
\vspmini
\end{figure}

An {\em isomorphism\/} between $\Sigma $-graphs $G,G':w$ is a graph isomorphism that
preserves the labeling information of the vertices. We shall not distinguish between
isomorphic graphs. Let $\calg _w(\Sigma )$ denote the set of $\Sigma $-graphs
of rank $w\in S^*$. The family $\calgsig =\{ \calg _w(\Sigma )\,|\,w\in S^*\}$ is
equipped with the indexed monoidal algebra operations (over the monoid $S^*$) 
as follows.
\vsp\newline
-- For a graph $G:w$, each permutation symbol $\rho :w\nnyil w'$ is interpreted as
the relabeling of the interfaces according to the $S$-permutation (not symbol!)
$\eps _{S^*}(\rho )$, where $\eps _{S^*}: (S^*)^*\nyil S^*$ is the counit map.
\newline
-- For graphs $G_1:w_1$ and $G_2:w_2$, $G_1\oplus G_2$ is the disjoint union of
$G_1$ and $G_2$ with the serial number of each interface vertex in $G_2$ incremented by
$|w_1|$.
\newline
-- The identity graph $\egy _w:ww$ for $w=A_1\ldots A_n$ is shown in Fig.~5c. The graph
$\egy _{()}$ is empty.
\newline
-- For a graph $G:wwv$ with $|w|=n$, the trace operation $\kapo _w$ is defined by gluing 
together the pairs of edges incident with the interface vertices having serial numbers 
$i$ and $n+i$ for each $1\leq i\leq n$, leaving out the interface vertices themselves. 
Whenever this procedure results in a loop of an even number of edges
(but no internal vertices) glued together,
a new loop vertex labeled by $\angle _A$ is created and added to the graph, where
$A$ is the common sort of the interface ports involved in the loop. 
\vsp

See \cite{hel} for
a more detailed description of $\kapo $ through examples. See also \cite{elg,blo,acta,tcs}
for the corresponding standard definition of feedback/iterarion in (directed) flow\-charts. 
Interestingly,
in all of these works, graphs (flowcharts) are equipped with a single loop vertex, so that
loops do not multiply when taking the feedback. On the other hand, the loop vertex is present in the 
graph $\egy _{()}$ as well. Regarding the single-sorted case
this amounts to imposing the additional axiom $\kapo \egy _1=\egy _0$ (rather, its
directed version, e.g.\ \cite[Axiom S5: $\uparrow 1=0$]{acta}), which is not
a standard traced monoidal category axiom. From the point of view of axiomatization this is a 
minor issue. Another issue, however, namely the assignment of an
individual monoid to each object $A$ is extremely important and interesting.
In terms of flowcharts,
this allows one to erase begin vertices and join two incoming edges at any given port. 
See e.g.\ the constants $0_1:0\nyil 1$ and $\eps:2\nyil 1$ in \cite {acta,tcs}. 
These constants (morphisms) were naturally incorporated in the axiomatization of schemes,
both flowchart and synchronous. Concerning undirected
graphs, the presence of such morphisms with a ``circularly symmetric'' interface allows for
an upgrade of ordinary edges to hyperedges, exactly the way it is described in
\cite{mil} for bigraphs. The axiomatization of undirected hypergraphs as SDCC categories 
will be presented in a forthcoming paper.

It is easy to check that the above interpretation of the indexed monoidal operations on
$\calg (\Sigma )$ satisfies the axioms I1-I9. Thus, $\calg (\Sigma )$ is an indexed
monoidal algebra over the monoid $S^*$. It is also clear that $\calg (\Sigma )$ is
generated by $\Sigma $, that is, by the collection of the atomic $\Sigma $-graphs.
(See again Fig.~5b.) Indeed, every undirected graph can be reconstructed from
its vertices as star graphs by adding internal edges one by one using the
trace operation. 
\begin{theorem}
The algebra $\calg (\Sigma )$ is freely generated by $\Sigma $. 
\end{theorem}
{\em Proof.\ \ }
Without essential loss of generality, we restrict our attention to the
single-sorted case. One way to prove the statement is to copy the normal form construction
for flowchart schemes as presented in \cite{acta}. Each step in this construction
\cite [Theorem 3.3]{mun} is completely analogous, except that one relies on undirected trace 
rather than directed feedback to create internal edges. 
Another idea that uses the corresponding result \cite[Corollary 2]{acta} 
directly
is the following. For each symbol $\sigma \in \Sigma _n$ ($n\geq 0$), consider the set of 
doubly ranked symbols $\sigma ^R_L$ of rank $(k,l)$ such that $L$ and $R$ are disjoint 
subsets of $[n]=\{ 1,\ldots ,n\}$ with cardinality $k$ and $l$ respectively, and
$k+l=n$. (Split the degrees into in-degrees and out-degrees in all possible ways.) Denote by
$\bar {\Sigma }$ the doubly ranked alphabet consisting of these new symbols. Construct
the free traced monoidal category $Sch(\bar {\Sigma })$  of $\bar {\Sigma }$-flowchart schemes
\cite{acta}, and consider the rank-preserving mapping $\Gamma :\sigma ^R_L\mapsto \sigma $ from
$\bar {\Sigma }$ into the SDCC category $\sdc \calg (\Sigma )$. Now let $\calm $ be an
arbitrary indexed monoidal algebra over monoid $M$ and specify $A\in M$ arbitrarily 
together with a mapping $\Omega : \sigma \mapsto f_{\sigma }$, where 
$f_{\sigma }: nA=A\otimes \ldots \otimes A$ is an arbitrary morphism in $\calm $.
Since $Sch(\bar {\Sigma })$ is freely generated by $\bar {\Sigma }$, there are unique
traced monoidal functors $\calf_1$ and $\calf_2$  from $Sch(\bar {\Sigma })$ into 
$\sdc \calg (\Sigma )$ and $\sdc \calm $ extending $\Gamma $ and $\Gamma \circ \Omega $,
respectively. One can then easily prove by induction that the desired unique SDCC functor
from $\sdc \calg (\Sigma )$ into $\sdc \calm $ extending $\Omega $ factors through an
arbitrary inverse of $\calf _1$, and $\calf _2$. Hence, the statement of the theorem
follows from Theorem~1.

A really elegant third proof, however, would use the $Int$ construction in \cite{tra}
-- alternatively, the $G$-construction in \cite{concur} -- by duplicating each
degree into an in-degree and a corresponding dual out-degree. The reader
familiar with either of these constructions will instantly recognize the point in this
argument. The statement of the theorem is, however, not an immediate consequence
of applying the construction to an appropriate traced monoidal category,
which is why we have chosen the above short and simple direct proof here.
\vege

Let $\calm $ be an arbitrary indexed monoidal algebra over $M$. An {\em interpretation\/}
of $\Sigma $ in $\calm $ is an alphabet mapping $\Omega : \Sigma \nyil \alp \calm $.
By Theorem~2, every interpration $\Omega $ can be extended in a unique way to a 
homomorphism $\bar {\Omega }:\calgsig \nyil \calm $. Thus, $\calg $ is indeed a
left adjoint for the functor $\alp $.
\section{Turing automata and Turing graph machines}
As an important example of indexed monoidal algebras, in this section we introduce the
algebra of Turing automata and Turing graph machines. We shall use the monoidal 
category $(\biset ,+)$ (small sets and bijections with disjoint union as tensor) as the
index category. Unfortunately, this category is not
strict, therefore the reader is asked to be lenient about the finer details. The only
``shaky'' ground will be the interpretation of set-permutation symbols as bijections
in the category $(\biset ,+)$, which is quite natural.
For a set $A$, let $A_*=A+\{ *\}$,
where $*$ is a fixed symbol, called the {\em anchor}. 
\vspmini
\begin{definition}
{\em A {\em Turing automaton\/} $T:A$ is a triple $(A,Q,\delta )$, where $A$ is a set of
{\em interfaces}, $Q$ is a nonempty set of {\em states}, and $\delta \subseteq (Q\times A_*)^2$
is the {\em transition relation}.}\vspmini
\end{definition}

The role of the anchor as a distinguished interface will be explained later.
The transition relation $\delta $ can either be considered as a function
$Q\times A_*\nyil \calp (Q\times A_*)$ or as a function $Q\times (A_*\times
A_*)\nyil \calp (Q)$, giving rise to a Mealy or Medvedev type automaton,
respectively. We shall favor the latter interpretation, and define $T$ to be
{\em deterministic\/} if $\delta $ is a partial function in this sense. Thus,
an input to automaton $T$ is a pair $(a,b)$ of interfaces. Nevertheless, we
still say that $T$ has a transition from $a$ to $b$ in state $q$, resulting in
state $r$, if $((q,a),(r,b))\in \delta $. By way of duality, one can also
consider $T$ as an automaton with states $A_*$ and inputs $Q\times Q$. We
shall reflect on this duality shortly. If $A$ is finite and $|A|=n$, then
$T$ can be viewed as an $(n+1)\times (n+1)$ matrix, where each entry is a
relation over $Q$. 
\vspp \newline 
{\bf Example}\ \ The
{\em $n$-ary atomic switch\/} is the Turing automaton $\cala _n:[n]$ ($n\geq 1$) 
having states $[n]$, so that  
\vspmmm
\[ \delta =\{ ((\ibf ,j),(\jbf ,i))\,|\, 1\leq i\neq j\leq n\}\cup 
\{ ((\ibf ,i), (\jbf ,j))\,|\, 1\leq i\neq j\leq n\} \vspmmm \]
\[
\cup
\{((\ibf ,*), (\ibf ,j)), ((\ibf ,j),(\ibf ,*)), ((\ibf ,*),(\ibf ,*))\,|\, i,j\in [n]\}. \]
For better readability, states, indicating a selected edge in an $n$-star graph,
 are written in boldface. In addition, if $n=1$, then
$((\egy , 1),(\egy , 1))\in \delta $. 

  Heuristically, the $n$-ary atomic switch captures the behavior of an atom in a molecule
having $n$ chemical bonds to neighboring atoms. Among these bonds exactly one is double,
and referred to as the {\em positive\/} edge in the underlying star graph. 
The mechanism of switching is then clear by the definition above. The active ingredient
(control) in this process is called the {\em soliton\/}, which is a form of energy traveling
in small packets through chains of alternating single and double bonds within the molecule,
causing the affected bonds to be flipped from single to double and vice versa.
See \cite{dav} for the physico-chemical details, and \cite {das,tcss,det} for the corresponding
mathematical model. Note that, by our definition above, whenever the soliton enters
an atom with a unique chemical bond (which must be double since $n=1$), it bounces
back immediately, producing no state change.
\vspp
 
  We now turn to defining the indexed monoidal algebra $\calt $ of Turing automata.
In this algebra, morphisms are Turing automata
$T:A$. A permutation symbol $\rho: A\nnyil B$ is interpreted as a relabeling of the
interfaces according to the unique bijection $A\nyil B$ determined by $\rho $ in $\biset $.
(Elaboration of details regarding the transition relation is left to the reader.)
The sum of $T:A$ and $T':B$ having states $Q$ and $Q'$, respectively, is the automaton
$T\oplus T'=(A+B, Q\times Q',\delta \oplus \delta ')$, where $\delta \oplus \delta '
\subseteq ((Q\times Q')\times (A+B)_*)^2$ is defined by 
\[ \mbox{$((q,q'),x),((r,r'),y))\in
\delta \oplus \delta '$ iff either $q'=r'$ and $((q,x),(r,y))\in \delta $, or
$q=r$ and $((q',x),(r',y))\in \delta '$.}\]
 (Notice the ambiguity in writing just
$x,y$ rather than $\langle x,A\rangle ,\langle y,A\rangle $ or $\langle x,B\rangle ,
\langle y,B\rangle $.) The definition, however, applies to
the case $x=*$ and/or $y=*$, too, so that taking the sum of $T$ and $T'$ amounts to
a selective performance of $\delta $ or $\delta '$ on $Q\times Q'$. 
The identity Turing automaton $\egy _A:A+A$ has a single state, in which there is a
transition from $\langle a,1\rangle $ to $\langle a,2\rangle $ and back for every
$a\in A$.

The definition of $\kapo _A T$ for a Turing automaton $T:A+A+B$ is complicated but
natural, and it holds the key to understanding the Turing-machine-like behavior of
this automaton. Intuitively, the definition models the behavior of loops in
flowchart algorithms \cite{elg} when implemented in an undirected environment.
That is, control enters $\kapo _A T$ at an interface $b\in B$, then, after alternating
between corresponding interfaces in $A+A$ any number of times, it leaves at another
(or the same) interface $b'\in B$. State changes are traced interactively during
this process. For technical reasons we shall restrict the formal definition of
trace to finitary Turing automata, whereby the number of interfaces is finite. This
will allow us to set up an analogy with the well-known Kleene construction
for converting a finite state automaton into a regular expression. 

Let $T=(A+A+B, Q,\delta )$ be a Turing automaton and $C\subseteq A$ be arbitrary 
such that $A=C+C'$. Following the Kleene construction, concentrate on the relations 
$\lambda _{x,y}^C\subseteq Q^2$ ($x,y\in (C'+B)_*$) as transitions of the automata
$\kapo _C T: C'+B$ from interface $x$ to interface $y$. Our goal is to satisfy the
Kleene formula
\vspmini
\[ \lambda _{x,y}^{C+\{ z\}}=\lambda _{x,y}^C\cup 
\left(
\begin{array}{cc}
\lambda ^C_{x,\langle z,1\rangle } & \lambda ^C_{x,\langle z,2\rangle }
\end{array}
\right)
\odot
\left(
\begin{array}{cc}
\lambda ^C_{\langle z,1\rangle ,\langle z,1\rangle } &
\lambda ^C_{\langle z,1\rangle ,\langle z,2\rangle } \\
\lambda ^C_{\langle z,2\rangle ,\langle z,1\rangle } &
\lambda ^C_{\langle z,2\rangle ,\langle z,2\rangle }
\end{array}
\right)
^{\star }
\odot 
\left(
\begin{array}{c}
\lambda ^C_{\langle z,1\rangle ,y} \\
\lambda ^C_{\langle z,2\rangle ,y}
\end{array} 
\right)
\vspmini \]
for every proper subset $C\subset A$ and $z\in A\setminus C$. In this
formula, $\odot $ denotes alternating matrix product in the sense
\vspmini
\[ U\odot 
\left(
\begin{array}{c}
V_1 \\ V_2
\end{array}
\right)
=
U\cdot
\left(
\begin{array}{c}
V_2 \\ V_1
\end{array}
\right)
,\vspmini \]
and $^{\star}$ is Kleene star of $2\times 2$ matrices based on $\odot $, that is,
$U^{\star }=\cup _{n\geq 0} U^n$, where $U^0$ is the alternate identity matrix 
$I_2\odot I_2$ and 
$U^{i+1}=U^i\odot U$. The underlying semiring $R$ is that of binary relations
over $Q$ with union as sum, composition as product, $id_Q$ as unit and $\emptyset $
as zero. Notice the immediate relationship between our $^{\star }$ and the star
operation in star theories as defined in \cite{iter}, e.g.\ in Conway matrix
theories. See also the Example in \cite{tra}, which originates from
\cite{iter}, too. As well, note the 
duality between states and input in comparison with the original Kleene formula.
The alternating matrix product is another hidden allusion to the $Int$ construction
\cite{tra} mentionned earlier, suggesting that the compact closed category resulting
from that construction be restricted to its ``self-dual'' objects $(A,A)$.

We are using the above formula as a recursive definition for $\lambda _{x,y}^C$, starting
from the basis step 
\[ \lambda _{x,y}^{\emptyset }=\{ (q,q')\,|\,((q,x),(q',y))\in \delta \}.\]
The transition relation $\hat {\delta }\subseteq (Q\times B_*)^2$ of $\kapo _A T$ is then set 
in such a way that $((q,b),(q',b'))\in \hat {\delta }$ iff $(q,q')\in \lambda _{b,b'}^A$.
In order to use this definition, one must prove that the specification of
$\lambda _{x,y}^C$ does not depend on the order in which the elements $z\in C'=A\setminus C$
are left out. This statement is essentially equivalent to axiom I9 (trace swapping).
\begin{theorem}
The algebra $\calt $ of Turing automata is indexed monoidal.
\end{theorem}
{\em Proof.\ \ }
At this point we can capitalize to a great extent on the simplicity of the indexed
monoidal algebra axioms. Indeed, each of these axioms, except for vanishing (I7)
and trace swapping (I9), holds naturally true in $\calt $. The vanishing axiom
expresses the fact that choosing the Kleene formula to define trace in $\calt $
is right, and trace
swapping ensures that the definition is correct. The proof of these two axioms
is left to the reader as an exercise. 
\vege

Finally, we explain the role of the anchor $*$. We did not want all Turing automata
of sort $I=\emptyset $ to have no transitions at all, like the automata $\angle _A=
\kapo _A \egy _A$, which all coincide, having a unique state. The
anchor is a fixed interface that is not supposed to be interconnected with any
other, so that automata in $\calt _I$ might still have transitions from $*$ to $*$.
The index category itself, however, need not be
that of pointed sets, because the anchor is not affected by any of the operations.

Let $D$ be a non-empty set of data. The indexed monoidal algebra $D$\dil $\calt $ of
$D$-{\em flow\/} Turing automata is defined in the following way.
\vsp\newline
-- Morphisms of sort $A$ are Turing automata $T:D\times A$.
\newline
-- Each permutation symbol $\rho :A\nnyil B$ is interpreted as a bijection (relabeling)
$D\times A\nyil D\times B$, which is basically $\rho _{\calt }$ performed on blocks
of size $D$ in parallel.
\newline
-- The operations sum and trace are adopted from $\calt $ (assuming the identification
of $D\times (A+B)$ with $D\times A+D\times B$), and
the identities $\egy _A$ are the identities $\egy _{D\times A}$ in $\calt $.
\vsp

  The notation $D$\dil\ originates from \cite{arnold}, where the magmoid
(single-sorted monoidal category) $k$\dil $\calm $ was introduced for integer
$k$ and magmoid $\calm $ along these lines. Intuitively, a $D$-flow Turing
automaton is a data-flow machine in which data in $D$ are passed along with
each transition. Notice that the anchor does not emit or receive any data. As
an immediate corollary to Theorem~3, the structure $D$\dil $\calt $ of
$D$-flow Turing automata is an indexed monoidal algebra. 
\begin{figure}[h]
\begin{center}
\includegraphics[scale=0.5]{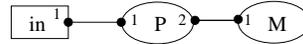}
\end{center}
\vspmini
\caption{The von Neumann machine}
\vspmini
\end{figure}

Consider, for
example, the scheme $N$ of the classical von Neumann computer in Fig.~6 as a data-flow
architecture. It consists of two interconnected single-sorted $D$-flow Turing automata: 
the processor $P:2$, and the memory $M:1$. The processor is a real finite state
automaton, having state components like registers, the instruction counter, the PSW, etc. 
The transitions of $P$ are very complex. On the other hand, $M$ has (practically) infinite states, 
but its transitions are straightforward. The set $D$ consists of all pieces of 
information (data, control, and/or address) that can be transmitted along the bus
line between $P$ and $M$ in either direction. The operation of $N$ need not be explained, 
and it is clearly that of a $D$-flow Turing automaton. It is a very
important observation, however, that the machine can do as much as we want in one
step, that is, from the time control enters port $1$ of $P$ until it leaves at
the same port. For example, it can execute one machine instruction stored in the
memory, or even a whole program stored there. In other words, semantics
is delay-free. In present-day digital computers this semantics is achieved by limiting
the scope of what the machine can do in one step through introducing clock
cycles and delay, which turn the computer into a synchronous system \cite{tcs}.
Theoretically speaking, undirected trace is turned into directed feedback with
delay (or, using an everyday language, recursion is transformed into a loop),
and  computations become inevitably directed in a rigid way. According to the original
scheme $N$, however, they need not be, yet they could be universal.
\vspp\newline
{\bf Example} (Continued)\ \ The $n$-ary atomic {\em alternating\/} switch 
$\cala ^2_n$ augments 
the ordinary $n$-ary atomic switch by the passing of a digital information in the following
way. Control from a negative interface (i.e., one not covered by the unique positive edge)
can only take 0 for input and emits 1 for output. (Remember that in the meantime the
positive edge is switched from the output side to the input side.) Conversely, control
from a positive interface can only take 1 for input and emits 0 for output. 
Transitions from and to the anchor are as in the corresponding $2n$-ary switch.

For the rest of the paper, the alphabet $\Sigma $ will be single-sorted, that is,
$\Sigma =\cup (\Sigma _n\,|\,n\geq 0)$.
\begin{definition} {\em A {\em $D$-flow Turing graph machine\/} 
over $\Sigma $ is a triple
$M=(G,D,\Omega )$, where $G$ is a $\Sigma $-graph and $\Omega $ is an interpretation of $\Sigma $
in $\calt $ under which the single sort of $\Sigma $ is mapped into the {\em finite\/} set $D$.
Equivalently, $\Omega $ is an interpretation in $D$\dil $\calt $ that maps sort
``1'' to object $\{1\}$.} 
\end{definition}

Intuitively, machine $M$ comes with an underlying graph $G$ that has a $D$-flow Turing 
automaton sitting in each of its internal vertices. The operation of $M$ as a complex
Turing automaton is uniquely determined by the given interpretation according to the homomorphism
$\bar \Omega $. The classical Turing machine concept is recaptured by taking $\Sigma =
\Sigma _2=\{c\}$, where $c$ stands for ``tape cell''. A Turing machine $TM$ is 
transformed into a $D$-flow Turing graph machine $M$ whose underlying graph is a linear
array of cell vertices with the following interpretation $T_c:2$ of $c$. The states of $T_c$ are
the tape symbols of $TM$, and, by way of duality, elements of $D$ are the states of $TM$.
The transition relation of $TM$ translates directly and naturally into that of $M$,
using duality. The only shortcoming of this
analogy is the finiteness of the underlying graph $G$, which can be ``compensated'' by
making the set of states $Q$ infinite, e.g., taking the colimit of finite approximation
automata in an appropriate extension of $\ima $ to a 2-category, whereby the
vertical structure is determined by homomorphisms of (Medvedev type) automata
in the standard sense.  Universality is then guaranteed either by the von Neumann machine
$N$ or by the universal Turing machine, as special Turing automata.

Returning to our dilemma of reversible vs.\ irreversible computations, we define the
``reverse'' of a Turing automaton $T=(A,Q,\delta )$ simply as $T^R=(A,Q,\delta ^{-1})$.
Although this definition is quite natural, one cannot expect that, for every computation
process represented by some Turing automaton $T$, both $T$ and $T^R$ be deterministic.
Indeed, this restriction would directly undermine universality. Eventually,
the point is not to actually perform the reverse of a given computation,
rather, being able to carry it out on a device that is in principle 
reversible. Turing graph machines do have this capability by definition.
Still, the effective construction of a universal Turing graph
machine remains an enormous challenge. The soliton automaton model described
below is an interesting try, but unfortunately it falls short of being universal
even in terms of designing individual ad-hoc machines.
To introduce this model as a Turing graph machine, let $\Sigma $ be the
ranked alphabet consisting of a single symbol $c_n$ for each rank $n\geq 1$.
\vspp\newline
{\bf Example} (Continued)\ \   A {\em pre-soliton automaton\/} is a Turing graph
machine $S=(G,\{0,1\},\Omega )$, where $\Omega $ is the fixed interpretation
that sends each symbol $c_n$ into the $n$-ary atomic alternating 
switch $\cala ^2_n$. 
Since the interpretation is fixed, we shall
identify each pre-soliton automaton with its underlying graph. Moreover, since
$\cala ^2_n$ is circularly symmetric, we do not need to
order the ports (degrees) of the internal vertices. Thus, $G$ is an ordinary undirected
graph (with its interface vertices still ordered, though). 
\vsp

Let $q$ be a state of graph $G$. By definition, each internal edge $e\in E$ is either {\em consistent\/}
with respect to $q$, meaning that $e$ has the same sign (positive or negative) viewed from its two
internal endpoints, or {\em inconsistent\/} if this is not the case. 
(Notice that a looping edge is always negative if consistent.) A {\em soliton walk\/}
from interface $i$ to interface $j$ ($i,j\in [n]+\{*\}$) is a transition of $G$ from $i$ to $j$
in state $q$ according to the standard behavior of $G$ as a Turing automaton. The 
reader can now easily verify that this definition of soliton walks coincides with the
original one given in \cite{das}, provided that $q$ is a {\em perfect internal matching\/} 
\cite {lov,tcss}
of $G$, that is, a state in which every edge of $G$ is consistent and the positive edges
determine a matching by which the internal vertices are all covered. Indeed, the definition
of $\cala ^2_n$ implies that the soliton can only traverse consistent
edges in an alternating positive-negative fashion.  A new feature of this model is a soliton
walk from the anchor, which must return to the anchor if $q$ is a perfect internal matching,
and in that case it defines a closed alternating walk (e.g.\ an alternating
cycle).

At this point we stop elaborating on soliton automata, leaving them as a subject for future
work. The key observation that allows one to restrict the states of pre-soliton automata to
perfect internal matchings is the Gallai-Edmonds Structure Theorem \cite{lov}, well-known in
matching theory. On the basis of this theorem, the Gallai-Edmonds algebra of graphs having
a perfect internal matching has been worked out in \cite{mun} as the homomorphic image of
the indexed monoidal algebra of graphs. The algebra (SDCC category) of soliton automata then turns out 
to be the quotient of $\calgsig $ determined by the pushout of the Gallai-Edmonds algebra homomorphism 
and $\bar \Omega $ in the category $\ima $. This result will be presented in a 
forthcoming paper. 
\bibliographystyle{eptcs} 

\end{document}